\documentclass{aip-cp}

\usepackage{graphicx} 
\usepackage{amssymb} 
\usepackage{amstext}
\usepackage{amsfonts} 
\usepackage{slashed} 
\usepackage[dvipsnames]{xcolor}

\newcommand{\nopieft}{\mbox{$\slashed{\pi}$EFT}} 
 
\newcommand{\Lag}{{\cal L}} 
\newcommand{\be}{\begin{equation}} 
\newcommand{\ee}{\end{equation}}

\def\lamb#1#2{$^{#1}_{\Lambda}${#2}} 
\def\lam#1#2{$^{#1}_{~\Lambda}${#2}} 
\def\la#1#2{$^{#1}_{~~\Lambda}${#2}} 

\begin{document}

\title{Old \& New in Strangeness Nuclear Physics} 
\author{Avraham Gal}
\affil{Racah Institute of Physics, The Hebrew University, 
91904 Jerusalem, Israel \\ avragal@savion.huji.ac.il}

\maketitle

\begin{abstract} 
Several persistent problems in strangeness nuclear physics are discussed in 
this opening talk at HYP2018, Norfolk VA, June 2018:~~(i) the \lamb{3}{H}, and 
\lamb{3}{n} if existing, lifetimes; (ii) charge symmetry breaking in $\Lambda$ 
hypernuclei; (iii) the overbinding of \lamb{5}{He} which might be related 
to the hyperon puzzle in neutron stars; and (iv) does $\Lambda^\ast$(1405) 
survive in strange hadronic matter? 
\end{abstract}

\section{INTRODUCTION} 
\label{sec:intro} 

\begin{figure}[htb] 
\includegraphics[width=0.48\textwidth,height=6.8cm]{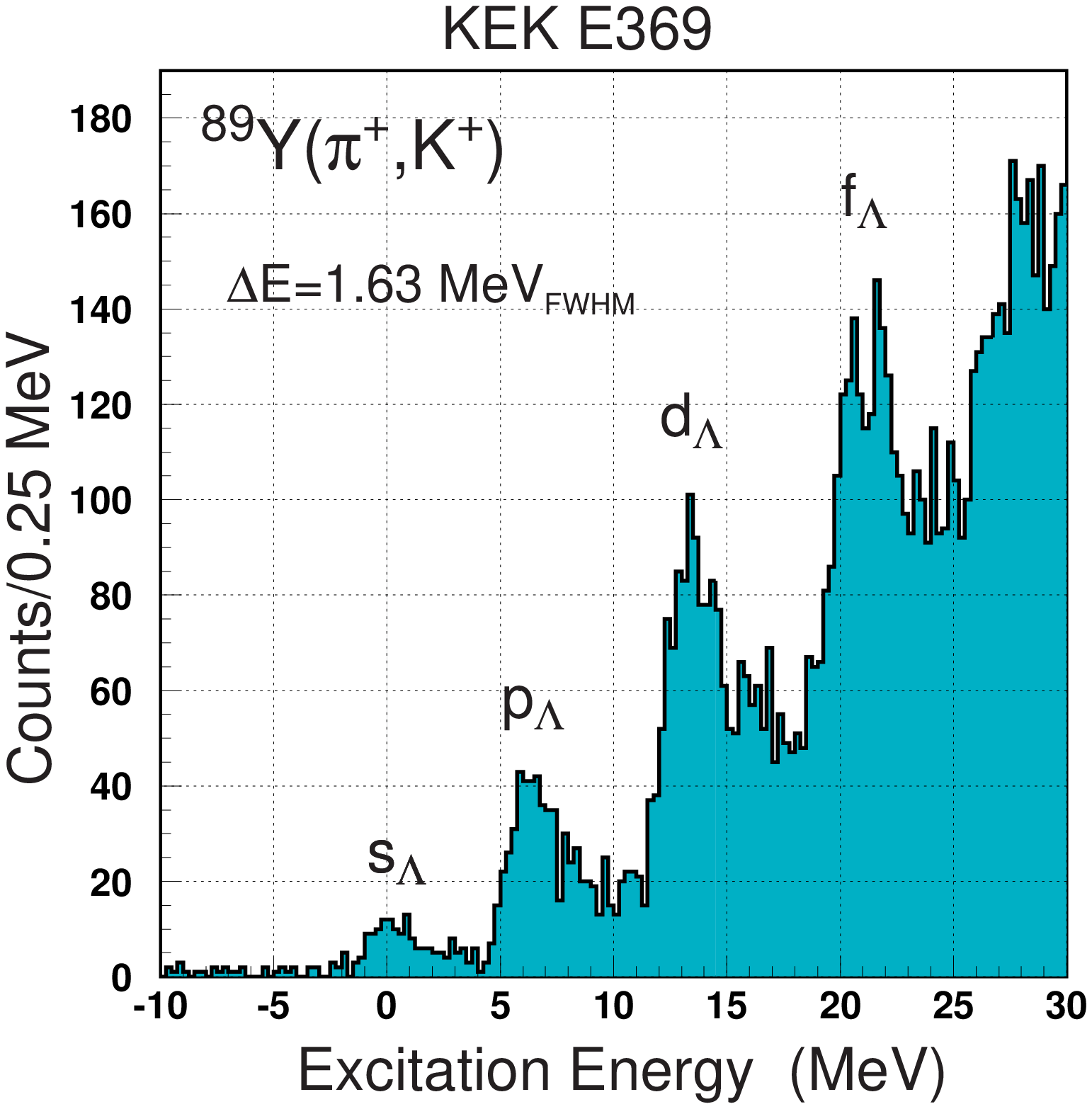}
\includegraphics[width=0.48\textwidth,height=6.8cm]{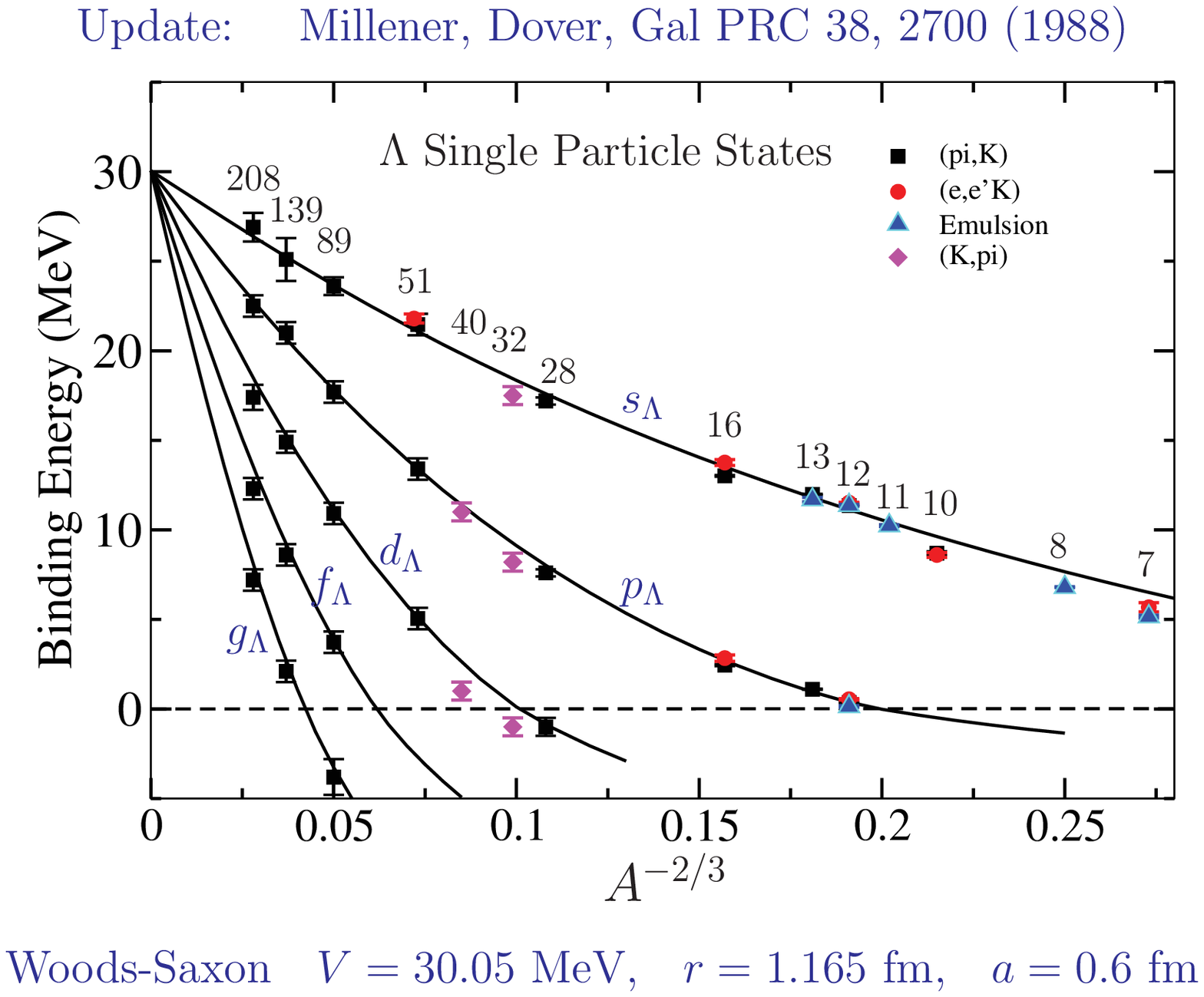}
\caption{Left: $^{89}$Y$(\pi^+,K^+)$ spectrum from KEK, exhibiting $\Lambda$ 
single-particle orbits~\cite{Hotchi01}. Right: compilation of $\Lambda$ 
binding energies in \lamb{7}{Li} to \la{208}{Pb} from different reactions, 
and as calculated using a 3-parameter WS potential~\cite{mdg88}. Figure 
adapted from Ref.~\cite{ghm16}.} 
\label{fig:mdg} 
\end{figure} 

Progress in Strangeness Nuclear Physics which 50 years ago consisted mostly of 
Hypernuclear Physics, with data collected exclusively in nuclear emulsions and 
bubble chambers, has been stepped up significantly with the advent of counter 
production experiments~\cite{ghm16}. The left panel of Fig.~\ref{fig:mdg} 
shows a $^{89}$Y$(\pi^+,K^+)$ spectrum from KEK~\cite{Hotchi01}, exhibiting 
distinct $\Lambda$ single-particle orbits in \lam{89}{Y} down to the 
ground-state (g.s.) $s_{\Lambda}$ orbit, a feature unparalleled in 
ordinary nuclei and hence termed ``a textbook example of a shell model 
at work"~\cite{mdg88}. 

The right panel of Fig.~\ref{fig:mdg} presents a compilation of most of 
the $\Lambda$ hypernuclear binding energies ($B_\Lambda$) measured across 
the periodic table as a function of $A^{-2/3}$ and as fitted by a simple 
3-parameter Woods-Saxon (WS) potential. The $\Lambda$ nuclear potential 
well depth $D_\Lambda$ in this simple-minded fit is 30 MeV, compared to 
27.8$\pm$0.3~MeV from the 1988 first theoretical analysis~\cite{mdg88} 
of the AGS $(\pi^+,K^+)$ data~\cite{Pile91}. Although the robustness of 
extrapolating to $A\to\infty$ is primarily owing to the $(\pi^+,K^+)$ 
production data in medium-weight and heavy nuclei, it is worth noting that 
$D_\Lambda$ was derived more than 50 years ago, to less accuracy of course, 
from $\pi^-$ decays of heavy spallation hypernuclei formed in silver and 
bromine emulsions~\cite{Lagnaux64,Lemonne65}:~~27$\pm$3~MeV (1964) and 
27.2$\pm$1.3~MeV (1965), respectively. 

\begin{figure}[htb] 
\includegraphics[width=0.48\textwidth]{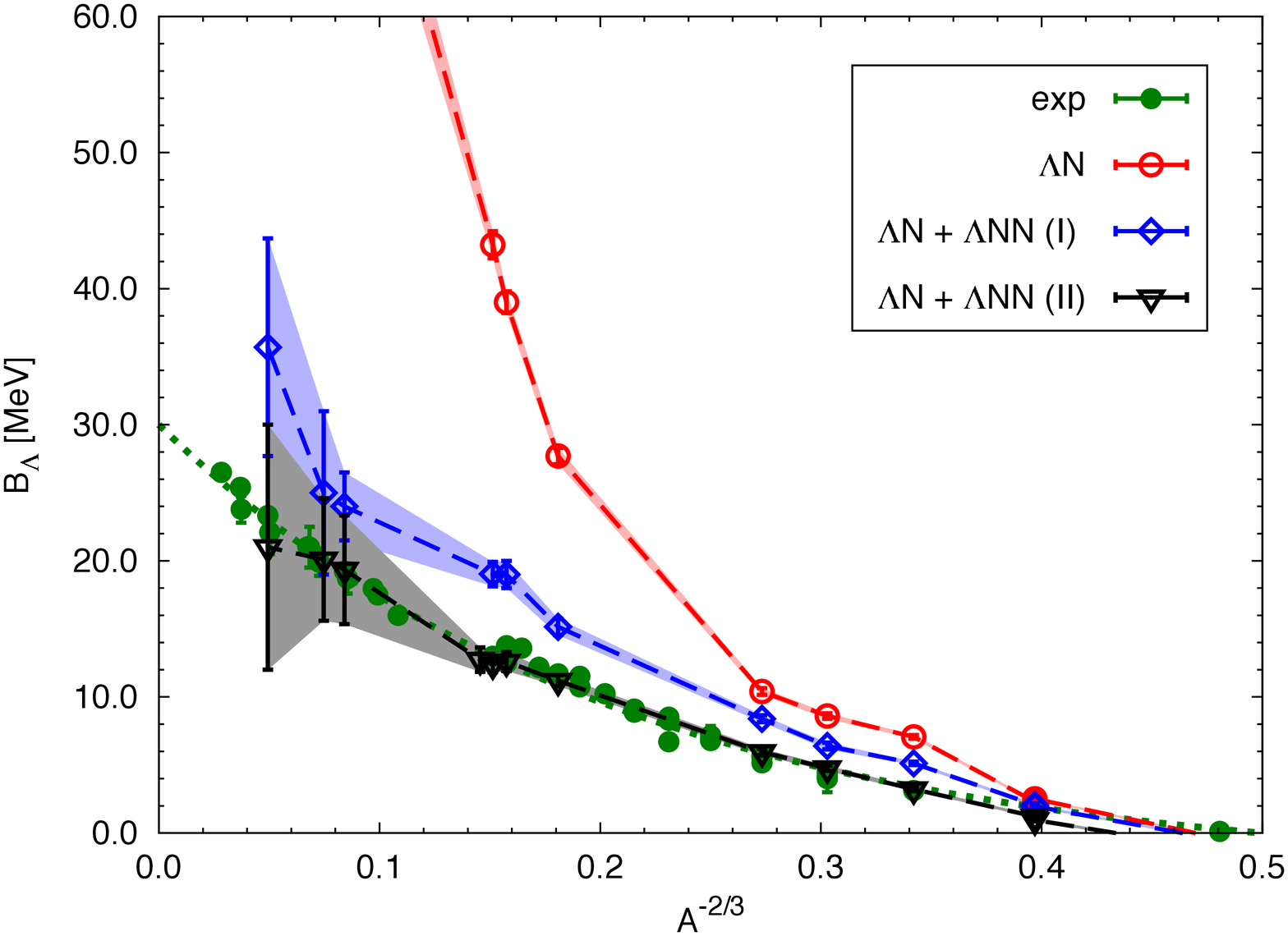}
\includegraphics[width=0.48\textwidth]{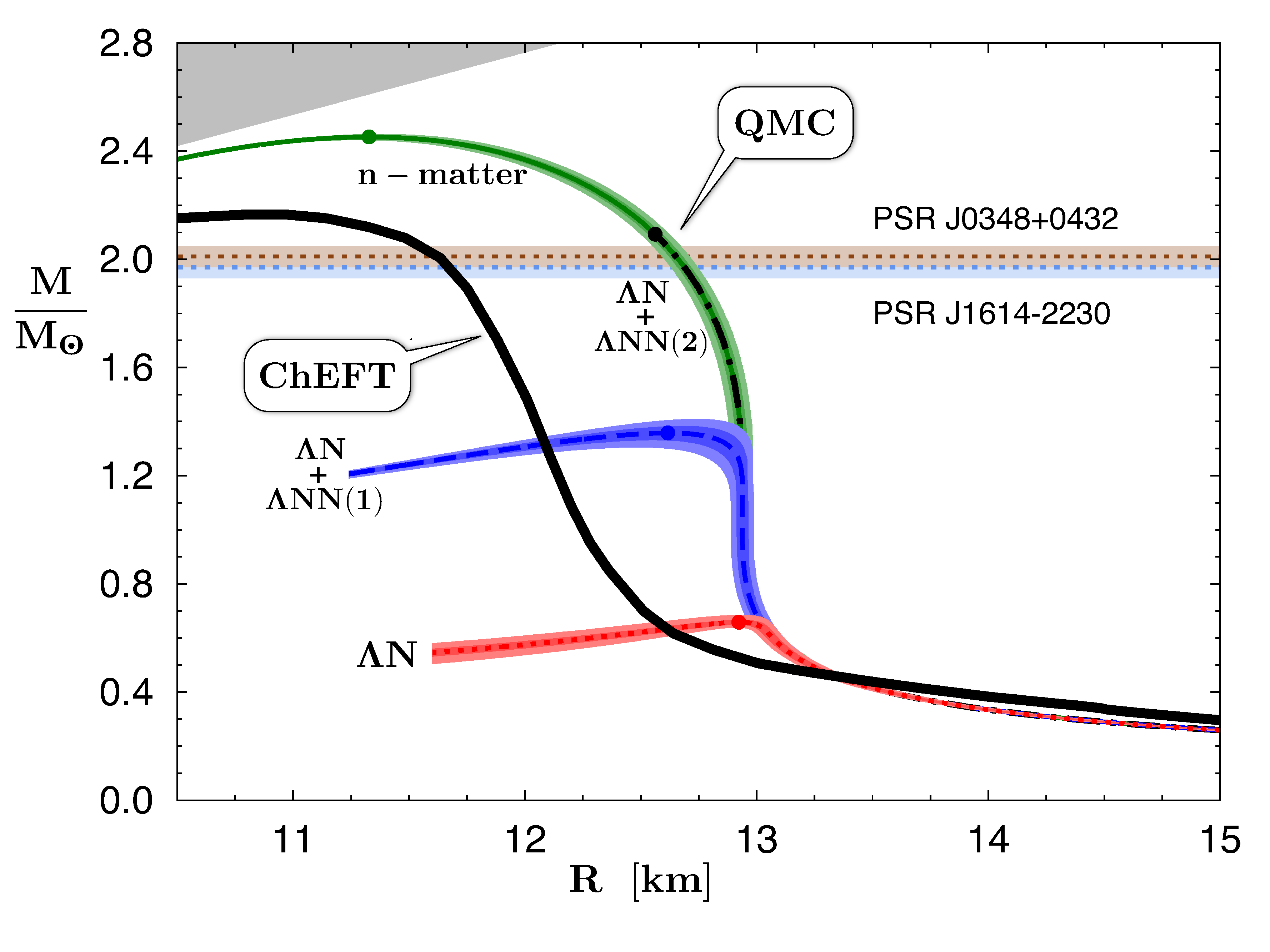}
\caption{Left: QMC calculations of hypernuclear g.s. $B_{\Lambda}$ values 
upon adding $\Lambda NN$ interactions I and II~\cite{lpg14}. Right: effect 
of adding $\Lambda NN$ repulsive interactions I and II in neutron star 
calculations~\cite{llgp15}. Figure adapted from Refs.~\cite{ghm16,weise15}.} 
\label{fig:lpg} 
\end{figure} 

A well-depth value of $D_\Lambda$$\sim$30~MeV cannot be reproduced using 
two-body $\Lambda N$ interactions exclusively. This was known already in 1988, 
with values about 50--60~MeV~\cite{mdg88} derived using reasonable $\Lambda N$ 
interactions, and is demonstrated perhaps in a somewhat exaggerated way in the 
left panel of Fig.~\ref{fig:lpg} taken from recent quantum Monte Carlo (QMC) 
calculations by Lonardoni {\it et al.}~\cite{lpg14}. These authors introduced 
phenomenological terms of repulsive $\Lambda NN$ interactions motivated by 
coupling $\Lambda N$ to $\Sigma N$ channels and back to $\Lambda N$ through 
one-pion exchange. The figure shows how by adding such terms and gradually 
strengthening their contribution (I to II), one is able to reproduce the 
empirical value of $D_\Lambda$$\sim$30~MeV. Implications of these results 
to neutron-star matter~\cite{llgp15} are demonstrated in the right panel of 
Fig.~\ref{fig:lpg}, showing neutron-star mass-radius curves corresponding 
to the three $B_{\Lambda}$ curves of the left panel. Adding $\Lambda NN$ 
repulsion stiffens the equation of state of neutron-star matter such that, 
for the choice II, $\Lambda$ hyperons are excluded and the experimentally 
deduced constraint $M/M_{\odot} > 2$ is satisfied. Shown for comparison 
is a curve (ChEFT) involving nucleons and pions only~\cite{weise15}. 
This apparent solution of the so called `hyperon puzzle' calls for more 
theoretically inclined work to confirm whether or not the puzzle has indeed 
been resolved. 

The overbinding of $\Lambda$ hypernuclei upon using exclusively two-body 
$\Lambda N$ interactions was realized in the early 1970s, with Dalitz 
{\it et al.}~\cite{dht72} demonstrating unambiguously that such overbinding 
persists already in \lamb{5}{He}, the lightest hypernucleus that is in fact 
a miniature of $\Lambda$ in nuclear matter. This old problem motivated the 
new work on $s$-shell hypernuclei outlined in the next section and expanded  
in a contribution by Contessi {\it et al.} in this Volume.  

Another topic discussed here, and separately by Andreas Nogga, is charge 
symmetry breaking (CSB) in $\Lambda$ hypernuclei as demonstrated by the 
emulsion large value 0.35$\pm$0.04~MeV~\cite{Davis05} of the binding energy 
difference $B_\Lambda$(\lamb{4}{He})$-$$B_\Lambda$(\lamb{4}{H}) which already 
in 1964 was realized to be as large, 0.30$\pm$0.14~MeV~\cite{Raymund64}. 
If a difference of two units in $N-Z$ induces this large binding energy 
difference, what should one expect going to $N-Z$ values as large as 44 in 
\la{208}{Pb}? The discussion of CSB in the $p$ shell, updating my HYP2015 
talk, partially answers this question.  

A third topic, presented by Benjamin D\"{o}nigus at HYP2018 and discussed 
here, is the hypertriton lifetime puzzle: why is the lifetime of \lamb{3}{H} 
so much shorter than the free $\Lambda$ lifetime, as suggested in recent 
measurements using relativistic heavy ion collisions to produce light nuclei, 
anti-nuclei, and in particular loosely-bound and spatially-extended 
hyperfragments such as \lamb{3}{H}? Also discussed briefly is the lifetime 
expected for the questionable \lamb{3}{n} hypernucleus. 

Finally, we discuss here briefly the role of the dynamically generated 
$\Lambda^\ast$(1405), a $\bar K N$ quasibound state, in strange hadronic 
matter. For a more expanded discussion see the contribution by 
Hrt\'{a}nkov\'{a} {\it et al.} in this Volume.

\section{OVERBINDING OF \lamb{5}{He}}
\label{sec:L5He} 

\begin{figure}[htb] 
\includegraphics[width=0.48\textwidth]{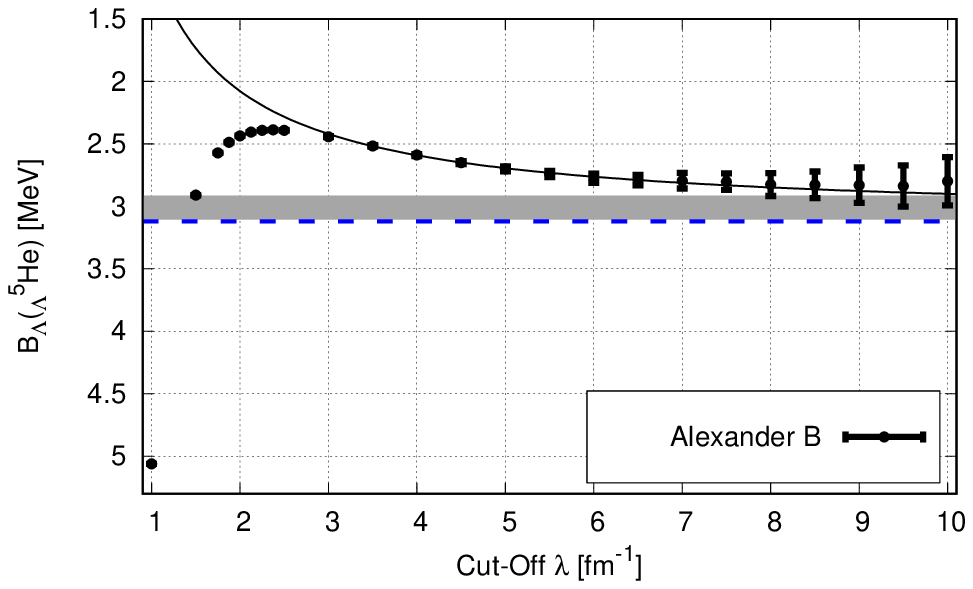}
\includegraphics[width=0.48\textwidth]{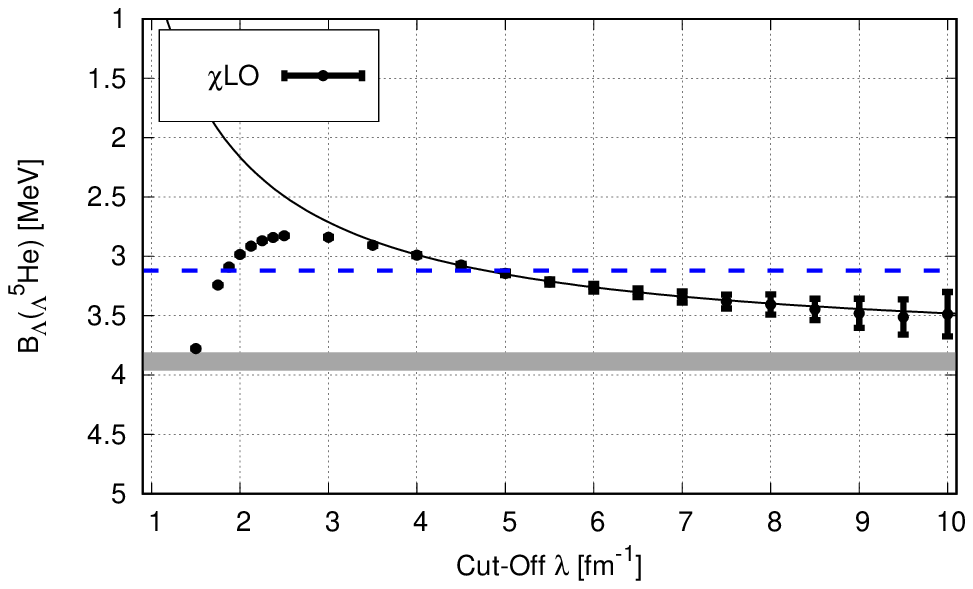}
\caption{$B_{\Lambda}$(\lamb{5}{He}) as a function of $\lambda$ in 
\nopieft~calculations with $\Lambda N$ input from $\Lambda p$ scattering 
experiments~\cite{Alex68} (left panel) and from a LO $\chi$EFT 
model~\cite{phm06} (right panel). Solid lines mark a fit $a+b/\lambda$ 
for $\lambda\geq 4$~fm$^{-1}$. Horizontal bands mark $\lambda\to\infty$ 
extrapolation uncertainties. Dashed horizontal lines mark the value 
$B_{\Lambda}^{\rm exp}(_{\Lambda}^5$He)=3.12$\pm$0.02~MeV. Figure adapted 
from Ref.~\cite{cbg18}.}  
\label{fig:cbg} 
\end{figure} 

The overbinding of \lamb{5}{He} upon using two-body $\Lambda N$ interactions, 
and also upon adding $\Lambda NN$ terms arising from the $\Lambda N$--$\Sigma 
N$ coupling, was first recognized and stated clearly in a 1972 landmark paper 
by Dalitz {\it et al.}~\cite{dht72}. There, as well as in version I of the QMC 
calculations~\cite{lpg14} (version II, instead, underbinds \lamb{3}{H} and 
\lamb{4}{H}~\cite{lgp13}) and in recent leading-order (LO) chiral effective 
field theory ($\chi$EFT) calculations~\cite{wr18}, the $\Lambda$ separation 
energy $B_{\Lambda}$(\lamb{5}{He}) comes out as large as 6~MeV, well above the 
measured value $B_{\Lambda}^{\rm exp}$(\lamb{5}{He})=3.12$\pm$0.02~MeV. 

Contessi {\it et al.} \cite{cbg18} revisited recently the overbinding problem 
by doing stochastic variational method (SVM) precise calculations of $s$-shell 
hypernuclei, using Lagrangians constructed at LO in a pionless effective field 
theory (\nopieft) approach limited to nucleons and $\Lambda$-hyperon degrees 
of freedom: 
\begin{equation} 
\Lag = N^\dagger (i\partial_0+\frac{\nabla^2}{2 M_N})\,N 
 + \Lambda^\dagger (i\partial_0+\frac{\nabla^2}{2 M_\Lambda})\,\Lambda 
 + \Lag_{2B}+\Lag_{3B}+\ldots\,,
\label{eq:Lag1} 
\end{equation} 
where $\Lag_{2B},\Lag_{3B},\ldots $ are two-body, three-body, etc., 
interaction terms composed of nucleon and $\Lambda$ fields and their 
derivatives subject to symmetry constraints that $\Lag$ is scalar and 
isoscalar and to a power counting which at LO limits these interaction terms 
to contact two-body and three-body $s$-wave interaction terms. These contact 
terms are regularized then by introducing a local Gaussian regulator with 
momentum cutoff $\lambda$. Apart from the two-body contact terms that are 
specified by $NN$ and $\Lambda N$ spin-singlet and triplet scattering lengths, 
amounting to four low-energy constants (LECs), the theory uses additionally 
four three-body LECs: a pure $NNN$ LEC fitted to $B(^3$H) and three $\Lambda 
NN$ LECs associated with the three possible $s$-wave $\Lambda NN$ systems, 
of which only \lamb{3}{H}($I$=0,\,$J^P$=${\frac{1}{2}}^+$) is bound. 
Therefore, on top of fitting its binding energy, the binding energies 
of \lamb{4}{H}$_{\rm g.s.}$($I$=${\frac{1}{2}}$,\,$J^P$=$0^+$) and of   
\lamb{4}{H}$_{\rm exc}$($I$=$\frac{1}{2}$,\,$J^P$=$1^+$) are also fitted. 
The fitted LECs are used then, for a sequence of $\lambda$ cutoff values, 
to evaluate the binding energies of $^4$He and \lamb{5}{He}. 

This \nopieft~approach was applied in SVM few-body calculations to the 
$s$-shell nuclei and in several models of the $\Lambda N$ scattering lengths 
also to the $s$-shell hypernuclei. The resulting $\Lambda$ separation energy 
values $B_{\Lambda}(_\Lambda^5$He) are shown in Fig.~\ref{fig:cbg} for two 
such models as a function of the cutoff $\lambda$. Common to all $\Lambda N$ 
models, the calculated $B_{\Lambda}(_\Lambda^5$He) values switch from about 
2--3~MeV overbinding at $\lambda$=1~fm$^{-1}$ to less than 1~MeV underbinding 
between $\lambda$=2 and 3~fm$^{-1}$, and smoothly varying beyond, approaching 
a finite (renormalization scale invariance) limit at $\lambda\to\infty$. 
A reasonable choice of {\it finite} cutoff values in the present case is 
between $\lambda$$\approx$1.5~fm$^{-1}$, which marks the \nopieft~breakup 
scale of 2$m_{\pi}$, and 4~fm$^{-1}$, beginning at which the detailed 
dynamics of vector-meson exchanges may require attention. 

The sign and size of the three-body contributions play a crucial role 
in understanding the cutoff $\lambda$ dependence of the calculated 
$B_{\Lambda}(_\Lambda^5$He). The nuclear $NNN$ term first changes from weak 
attraction at $\lambda$=1~fm$^{-1}$ in $^3$H and $^4$He, similar to that 
required in phenomenological models~\cite{nkg00}, to strong repulsion at 
$\lambda$=2~fm$^{-1}$, which reaches maximal values around $\lambda$=4~fm$^{-1}
$. However, for larger values of $\lambda$ it decreases slowly. The $\Lambda 
NN$ contribution follows a similar trend, but it is weaker than the $NNN$ 
contribution by a factor of roughly 3 when repulsive. The transition of the 
three-body contributions from long-range weak attraction to relatively strong 
repulsion for short-range interactions is correlated with the transition seen 
in Fig.~\ref{fig:cbg} from strongly overbinding $^5_\Lambda$He to weakly 
underbinding it. We note that for $\lambda\gtrsim 1.5$~fm$^{-1}$ all of the 
three $\Lambda NN$ components are repulsive, as required to avoid Thomas 
collapse, imposing thereby some constraints on the $\Lambda NN$ LECs. 

Momentum-dependent interaction terms, such as tensor and spin-orbit, which 
appear at subleading order in $\nopieft$ power counting, will have to be 
introduced in future applications to $p$-shell hypernuclei. The long-range 
$\Lambda N$ tensor force induced by a $\Lambda N\to\Sigma N$ one-pion exchange 
(OPE) transition followed by a $\Sigma N\to\Lambda N$ OPE transition is 
expected to be weak because this two-pion exchange mechanism is dominated 
by its central $S\to D\to S$ component, which is partially absorbed in the 
$\Lambda N$ and $\Lambda NN$ LO contact LECs. Short-range $K$ and $K^\ast$ 
meson exchanges induce a mild $\Lambda N$ tensor force~\cite{gsd71,mgdd85}, 
the weakness of which is confirmed in shell-model studies of observed 
$p$-shell $\Lambda$ hypernuclear spectra~\cite{mil12}. Recent LO $\chi$EFT 
calculations~\cite{wr16} using induced $YNN$ repulsive contributions 
suggest that the $s$-shell overbinding problem extends to the $p$ shell. 
In contrast, shell-model studies~\cite{mil12} reproduce satisfactorily 
$p$-shell ground-state $B_{\Lambda}$ values, essentially by using 
$B_{\Lambda}^{\rm exp}(_{\Lambda}^5$He) for input, except for the relatively 
large difference of about 1.8~MeV between $B_{\Lambda}({_{\Lambda}^9}$Li) 
and $B_{\Lambda}({_{\Lambda}^9}$Be). In fact, it was noted long ago 
that strongly repulsive $\Lambda NN$ terms could settle it~\cite{gal67}. 
It would be interesting to apply our derived $\Lambda NN$ interaction 
terms in future shell-model calculations. 

Finally, the \nopieft~Lagrangian used here includes already at LO repulsive 
$\Lambda NN$ terms which are qualitatively as strong as those used by 
Lonardoni, Pederiva and Gandolfi~\cite{lpg14} to resolve the hyperon puzzle 
\cite{llgp15}. It would be interesting then to apply our $\Lambda N$+$\Lambda 
NN$ interaction terms in state-of-the-art neutron-star matter calculations 
to see whether or not their suggested resolution of the hyperon puzzle is 
sufficiently robust.

\section{CHARGE SYMMETRY BREAKING} 
\label{sec:CSB} 

\begin{figure}[hbt] 
\includegraphics[width=0.48\textwidth,height=5cm]{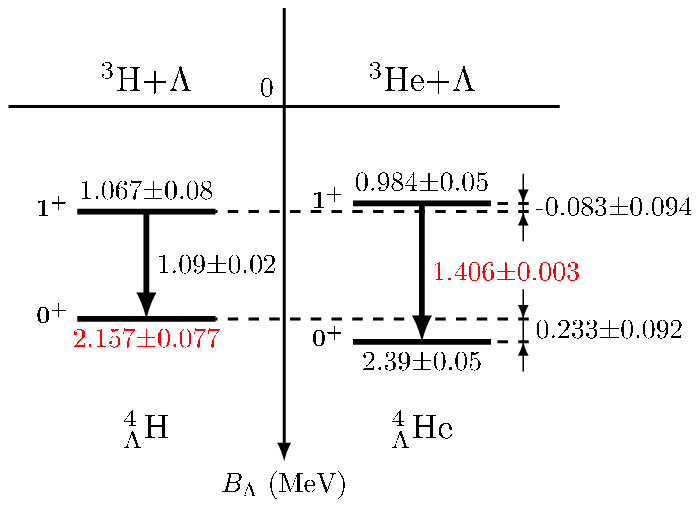} 
\includegraphics[width=0.48\textwidth,height=5cm]{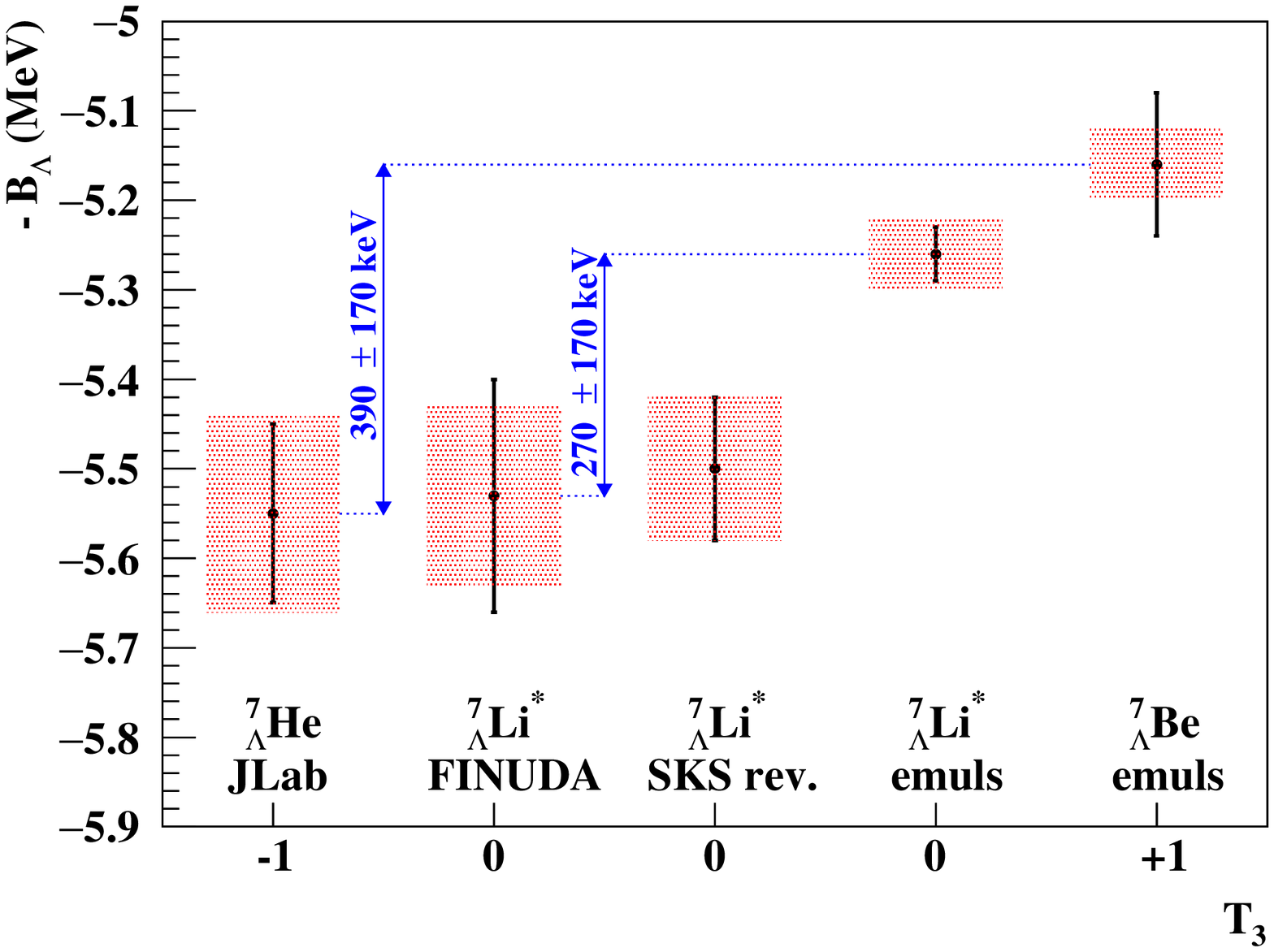} 
\caption{Left: $A=4$ hypernuclear level diagram. Recent determinations of the 
\lamb{4}{He} excitation energy $E_{\gamma}(1^+_{\rm exc}\to 0^+_{\rm g.s.})$ 
from J-PARC~\cite{E13} and of the \lamb{4}{H} $0^+_{\rm g.s.}$ binding energy 
from MAMI~\cite{MAMI15,MAMI16} are marked in red. CSB splittings are shown to 
the right of the \lamb{4}{He} levels; figure adapted from Ref.~\cite{MAMI16}. 
Right: $A=7$ hypernuclear $J^{\pi}={\frac{1}{2}}^+$ $I=1$ level diagram 
adapted from Ref.~\cite{botta17}. CSB splittings are relatively small if 
derived from {\it either} counter experiments {\it or} emulsion exposures.} 
\label{fig:A=4} 
\end{figure}

Charge symmetry breaking (CSB) is particularly strong in the $A=4$ mirror 
hypernuclei \lamb{4}{H}--\lamb{4}{He}, as shown in Fig.~\ref{fig:A=4}. 
Although OPE does not contribute directly to the $\Lambda N$ strong 
interaction owing to isospin invariance, it does contribute as pointed out 
by Dalitz and Von Hippel (DvH) through a CSB potential $V_{\rm CSB}^{\rm OPE}$ 
generated by admixing the SU(3) octet $\Lambda_{I=0}$ and $\Sigma^0_{I=1}$ 
hyperons in the physical $\Lambda$ hyperon~\cite{DvH64}. For the mirror 
\lamb{4}{H}--\lamb{4}{He} ground-state (g.s.) levels built on the $^3$H--$^3
$He cores, and using the DvH purely central wavefunction, the OPE CSB 
contribution amounts to $\Delta B_{\Lambda}^{J=0}$$\approx$95~keV where 
$\Delta B_{\Lambda}^J$$\equiv$$ B_{\Lambda}^J$(\lamb{4}{He})$-$$B_{\Lambda}^J
$(\lamb{4}{H}). This was confirmed recently by Gazda and Gal using $A$=4 
wavefunctions generated within a LO $\chi$EFT no-core shell-model (NCSM) 
calculation~\cite{gg16a} in which the OPE tensor interaction adds 
$\approx$100~keV~\cite{gg16b}. Remarkably, the OPE overall contribution 
of $\approx$200~keV to the CSB splitting of the \lamb{4}{H}--\lamb{4}{He} 
mirror g.s. levels roughly agrees with the large observed g.s. CSB splitting 
$\Delta B^{J=0}_{\Lambda}$=233$\pm$92~keV shown in Fig.~\ref{fig:A=4} which 
is considerably larger than the approximately 70 keV CSB part of the 764~keV 
Coulomb-dominated binding-energy difference of the $^3$H--$^3$He core mirror 
nuclei, driven apparently via short-range $\rho$-$\omega$ mixing. 

In addition to OPE, $\Lambda-\Sigma^0$ mixing affects also short-range meson 
exchanges (e.g., vector mesons) that in $\chi$EFT are replaced by contact 
terms. Quite generally, in baryon-baryon models that include {\it explicitly} 
a charge-symmetric (CS) $\Lambda N\leftrightarrow\Sigma N$ ($\Lambda\Sigma$) 
coupling, the direct $\Lambda N$ matrix element of $V_{\rm CSB}$ is obtained 
from a strong-interaction CS $\Lambda\Sigma$ coupling matrix element 
$\langle N\Sigma|V_{\rm CS}|N\Lambda\rangle$ by 
\begin{equation} 
\langle N\Lambda|V_{\rm CSB}|N\Lambda\rangle = -0.0297\,\tau_{Nz}\,\frac{1}
{\sqrt{3}}\,\langle N\Sigma|V_{\rm CS}|N\Lambda\rangle , 
\label{eq:OME} 
\end{equation} 
where the $z$ component of the nucleon isospin Pauli matrix ${\vec\tau}_N$ 
assumes the values $\tau_{Nz}=\pm 1$ for protons and neutrons, respectively, 
the isospin Clebsch-Gordan coefficient $1/\sqrt{3}$ accounts for the 
$N\Sigma^0$ amplitude in the $I_{NY}=1/2$ $N\Sigma$ state, and the space-spin 
structure of this $N\Sigma$ state is taken identical to that of the $N\Lambda$ 
state sandwiching $V_{\rm CSB}$. The 3\% CSB scale factor $-$0.0297 in 
Eq.~(\ref{eq:OME}) follows by evaluating the $\Lambda-\Sigma^0$ mass mixing 
matrix element $\langle\Sigma^0|\delta M|\Lambda\rangle$ from SU(3) mass 
formulae \cite{DvH64,gal15}. 

\begin{figure}[!t]
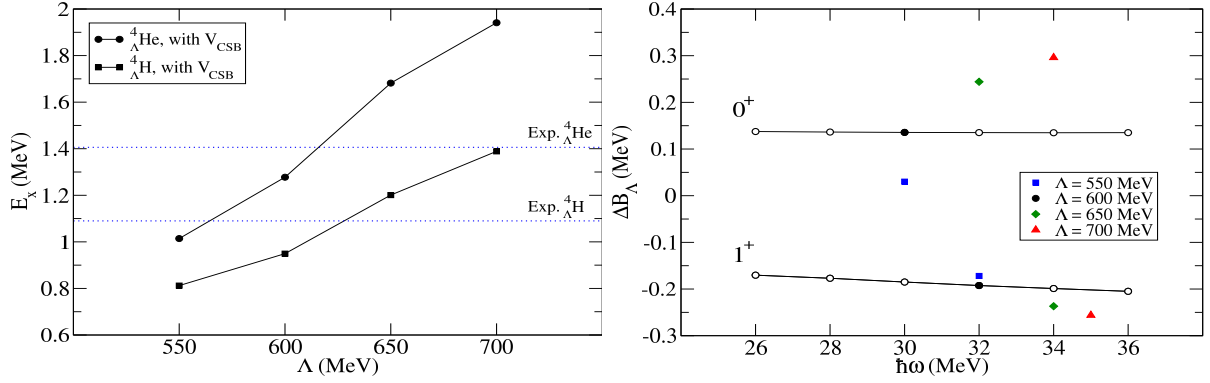
  
\includegraphics[width=0.48\textwidth,height=5cm]{ex-l_vm_csb.eps} 
\includegraphics[width=0.48\textwidth,height=5cm]{csb-om_vm.eps} 
\caption{NCSM calculations of \lamb{4}{H} and \lamb{4}{He}, using CS LO 
$\chi$EFT $YN$ interactions~\cite{phm06} and $V_{\rm CSB}$, Eq.~(\ref{eq:OME}), 
derived from these CS interactions. Left: momentum cutoff dependence of 
excitation energies $E_{\rm x}$(0$^+_{\rm g.s.}$$\to$1$^+_{\rm exc}$). 
The $\gamma$-ray measured values of $E_{\rm x}$ from Fig.~\ref{fig:A=4} are 
marked by dotted horizontal lines. Right: HO $\hbar\omega$ dependence, for 
$\Lambda$=600~MeV, of the separation-energy differences $\Delta B_{\Lambda}^J$ 
for $0^+_{\rm g.s.}$ (upper curve) and for $1^+_{\rm exc}$ (lower curve). 
Results for other values of $\Lambda$ are shown at the respective absolute 
variational energy minima. Figure adapted from Ref.~\cite{gg16b}.} 
\label{fig:csb} 
\end{figure} 

Since the CSB $\Lambda N$ matrix element in Eq.~(\ref{eq:OME}) is given in 
terms of strong-interaction CS $\Lambda\Sigma$ coupling, it is interesting to 
see how strong the latter is in realistic microscopic $YN$ interaction models. 
Recent four-body calculations of $_{\Lambda}^4{\rm He}$ levels \cite{gazda14}, 
using the Bonn-J\"{u}lich LO $\chi$EFT SU(3)-based $YN$ CS potential 
model~\cite{phm06}, show that almost 40\% of the $0^+_{\rm g.s.}\to 1^+_{
\rm exc}$ excitation energy $E_x$ arises from $\Lambda\Sigma$ coupling. 
This also occurs in the NSC97 models~\cite{NSC97} as demonstrated by Akaishi 
{\it et al.}~\cite{akaishi00}. With $\Lambda\Sigma$ matrix elements of 
order 10~MeV, the 3\% CSB scale factor in Eq.~(\ref{eq:OME}) suggests 
a CSB splitting $\Delta E_{\rm x}\equiv E_{\rm x}$(\lamb{4}{He})$-E_{\rm x}
$(\lamb{4}{H})$\sim$300~keV, in good agreement with the splitting 
$\Delta E_{\rm x}^{\rm exp}=320\pm 20$~keV~\cite{E13} deduced from 
the r.h.s. of Fig.~\ref{fig:A=4}. 

Results of the Gazda-Gal four-body NCSM calculations of the $A$=4 hypernuclei 
\cite{gg16a,gg16b}, using the Bonn-J\"{u}lich model \cite{phm06} with momentum 
cutoff in the range $\Lambda$=550--700~MeV, are shown in Fig.~\ref{fig:csb}. 
Plotted on the l.h.s. are the calculated $0^+_{\rm g.s.}\to 1^{+}_{\rm exc}$ 
excitation energies in \lamb{4}{H} and in \lamb{4}{He}, both of which are 
found to increase with $\Lambda$ such that somewhere between $\Lambda$=600 
and 650 MeV the $\gamma$-ray measured values of $E_{\rm x}$ are reproduced. 
The $\Lambda-\Sigma^0$ mixing CSB splitting $\Delta E_x$ obtained by using 
Eq.~(\ref{eq:OME}) also increases with $\Lambda$ such that for $\Lambda
$=600~MeV the calculated value $\Delta (\Delta B_{\Lambda})\equiv\Delta 
B_{\Lambda}^{\rm calc}(0^+_{\rm g.s.})-\Delta B_{\Lambda}^{\rm calc}(1^+_{
\rm exc})=330\pm 40$~keV derived from the r.h.s. of the figure agrees with 
$\Delta E_{\rm x}^{\rm exp}$. 

Plotted on the r.h.s. of Fig.~\ref{fig:csb} is the $\hbar\omega$ dependence 
of $\Delta B^{J}_{\Lambda}$, including $V_{\rm CSB}$ from Eq.~(\ref{eq:OME}) 
and using $N_{\rm max}\to\infty$ extrapolated values for each of the four 
possible $B^{J}_{\Lambda}$ values calculated at cutoff $\Lambda$=600~MeV. 
Extrapolation uncertainties for $\Delta B^{J}_{\Lambda}$ are 10 to 20~keV. 
$\Delta B^{J=0}_{\Lambda}$ varies over the spanned $\hbar\omega$ range by 
a few keV, whereas $\Delta B^{J=1}_{\Lambda}$ varies by up to $\sim$30~keV. 
Figure~\ref{fig:csb} demonstrates a strong (moderate) cutoff dependence of 
$\Delta B^{J=0}_{\Lambda}$ ($\Delta B^{J=1}_{\Lambda}$): 
\begin{equation} 
\Delta B^{J=0}_{\Lambda}=177^{+119}_{-147}~{\rm keV},\,\,\,\,\,\, 
\Delta B^{J=1}_{\Lambda}=-215^{+43}_{-41}~{\rm keV}. 
\label{eq:DelB} 
\end{equation} 
The opposite signs and roughly equal sizes of these $\Delta B^J_{\Lambda}$ 
values follow from the dominance of the $^1S_0$ contact term (CT) in the 
$\Lambda\Sigma$ coupling potential of the LO $\chi$EFT $YN$ Bonn-J\"{u}lich 
model~\cite{phm06}, whereas the PS SU(3)-flavor octet (${\bf 8_{\rm f}}$) 
meson-exchange contributions are relatively small and of opposite sign to 
that of the $^1S_0$ CT contribution. This paradox is resolved by noting that 
regularized pieces of Dirac $\delta(\bf r)$ potentials that are discarded in 
the classical DvH treatment survive in the LO $\chi$EFT PS meson-exchange 
potentials. Suppressing such a zero-range regulated piece of CSB OPE within 
the full LO $\chi$EFT $A$=4 hypernuclear wavefunctions gives~\cite{gg16b} 
\begin{equation} 
{\rm OPE (DvH):}\,\,\,\,\,\,\Delta B^{J=0}_{\Lambda}\approx 
175\pm 40~{\rm keV},\,\,\,\,\,\, \Delta B^{J=1}_{\Lambda}\approx 
-50\pm 10~{\rm keV},  
\label{eq:OPE} 
\end{equation}  
with smaller momentum cutoff dependence uncertainties than in 
Eq.~(\ref{eq:DelB}). Both Eqs.~(\ref{eq:DelB}) and (\ref{eq:OPE}) agree 
within uncertainties with the CSB splittings $\Delta B^J_{\Lambda}$ marked 
in Fig.~\ref{fig:A=4}.  

To apply the CSB procedure defined by Eq.~(\ref{eq:OME}) to $p$-shell 
hypernuclei, one introduces effective CS $\Lambda\Sigma$ central interactions 
with $p$-shell $0p_N0s_Y$ matrix elements ${\bar V}^{0p}_{\Lambda\Sigma}$ and 
$\Delta^{0p}_{\Lambda\Sigma}$ listed in the caption to Table~\ref{tab:pshell}. 
These matrix elements follow from the shell-model reproduction of hypernuclear 
$\gamma$-ray transition energies by Millener~\cite{mil12} and are smaller by 
a factor roughly two than the corresponding $s$-shell $0s_N0s_{Y}$ matrix 
elements, thereby resulting in smaller $\Sigma$ hypernuclear admixtures and 
implying that CSB contributions in the $p$ shell are weaker with respect 
to those in the $A=4$ hypernuclei also by a factor of two, as demonstrated 
in Fig.~\ref{fig:A=4}. To evaluate $p$-shell CSB contributions, the 
single-nucleon expression (\ref{eq:OME}) is extended by summing over 
$p$-shell nucleons~\cite{gal15}:  
\begin{equation} 
V_{\rm CSB} = -0.0297\,\frac{1}{\sqrt{3}}
\sum_j{({\bar V}^{0p}_{\Lambda\Sigma}+\Delta^{0p}_{\Lambda\Sigma}\,
{\vec s}_j\cdot{\vec s}_Y)\,\tau_{jz}}. 
\label{eq:VCSB} 
\end{equation} 
Results of applying this effective $\Lambda\Sigma$ coupling model to several 
pairs of g.s. levels in $p$-shell hypernuclear isomultiplets are given in 
Table~\ref{tab:pshell}, adapted from Ref.~\cite{gg18}. All pairs except for 
$A=7$ are g.s. mirror hypernuclei identified in emulsion~\cite{Davis05} where 
binding energy systematic uncertainties are largely canceled out in forming 
the listed $\Delta B_{\Lambda}^{\rm exp}$ values. In the case of the 
(\lamb{7}{He},~\lamb{7}{Li}$^{\ast}$,~\lamb{7}{Be}) isotriplet ${\frac{1}{2}}^
+$ levels shown in Fig.~\ref{fig:A=4}, whereas emulsion $B_{\Lambda}^{\rm exp}
$(g.s.) values  supplemented by the observation of a 3.88~MeV $\gamma$-ray 
transition \lamb{7}{Li}$^{\ast}\to\gamma$+\lamb{7}{Li}~\cite{tamura00} 
were used for the \lamb{7}{Be}--\lamb{7}{Li}$^{\ast}$ pair, recent counter 
measurements that provide absolute energy calibrations relative to precise 
values of free-space known masses were used for the \lamb{7}{Li}$^{\ast}
$--\lamb{7}{He} pair~\cite{botta17} (FINUDA for \lamb{7}{Li}$_{\rm g.s.}$ 
$\pi^-$ decay~\cite{botta09} and JLab for \lamb{7}{He} 
electroproduction~\cite{JLabL7He}). Note that the value reported by FINUDA 
for $B_{\Lambda}$(\lamb{7}{Li}$_{\rm g.s.}$), 5.85$\pm$0.17~MeV, differs from 
the emulsion value of 5.58$\pm$0.05~MeV. Recent $B_{\Lambda}$ values from JLab 
electroproduction experiments for \lamb{9}{Li}~\cite{JLabL9Li} and 
\lam{10}{Be}~\cite{JLabL10Be} were not used for lack of similar data 
on their mirror partners. 

\begin{table}[htb] 
\caption{$\langle V_{\rm CSB}\rangle$ contributions (in keV) to $\Delta 
B^{\rm calc}_{\Lambda}$ in $p$-shell hypernuclei g.s. isomultiplets, 
using $\Lambda\Sigma$ coupling matrix elements ${\bar V}^{0p}_{\Lambda\Sigma}
$=1.45~MeV and $\Delta^{0p}_{\Lambda\Sigma}$=3.04~MeV~\cite{mil12} in 
Eq.~(\ref{eq:VCSB}). A similar calculation for the $s$-shell $A$=4 mirror 
hypernuclei~\cite{gal15} is included for comparison. Listed values of 
$\Delta B_{\Lambda}^{\rm exp}$ are based on g.s. emulsion data~\cite{Davis05} 
except for \lamb{4}{He}--\lamb{4}{H} \cite{MAMI16} and \lamb{7}{Li}$^{\ast}-
$\lamb{7}{He} \cite{botta17}.} 
\vspace{5pt} 
{\renewcommand{\arraystretch}{1.1}} 
\begin{tabular}{ccccccc} 
\hline 
\lamb{A}{Z$_{>}$}--\lamb{A}{Z$_{<}$} & \lamb{4}{He}--\lamb{4}{H} & 
\lamb{7}{Be}--\lamb{7}{Li}$^{\ast}$ & \lamb{7}{Li}$^{\ast}$--\lamb{7}{He} & 
\lamb{8}{Be}--\lamb{8}{Li} & \lamb{9}{B}--\lamb{9}{Li} & 
\lam{10}{B}--\lam{10}{Be} \\ 
$I,J^{\pi}$ & $\frac{1}{2},0^+$ & $1,{\frac{1}{2}}^+$ & $1,{\frac{1}{2}}^+$ & 
$\frac{1}{2},1^-$ & $1,{\frac{3}{2}}^+$ & $\frac{1}{2},1^-$ \\ 
\hline
$\langle V_{\rm CSB}\rangle$ & 232 & 50 & 50 & 119 & 81 & 17 \\ 
$\Delta B_{\Lambda}^{\rm calc}$ & 226 & $-$17& $-$28 & $+$49 & $-$54 & 
$-$136 \\ 
$\Delta B_{\Lambda}^{\rm exp}$ & 233$\pm$92 & $-$100$\pm$90 & 
$-$20$\pm$230 & $+$40$\pm$60 & $-$210$\pm$220 & $-$220$\pm$250 \\ 
\hline
\end{tabular} 
\label{tab:pshell} 
\end{table}

The listed $A=7-10$ values of 
$\langle V_{\rm CSB}\rangle$ exhibit strong SU(4) correlations, highlighted 
by the enhanced value of 119~keV for the SU(4) nucleon-hole configuration 
in \lamb{8}{Be}--\lamb{8}{Li} with respect to the modest value of 17~keV 
for the SU(4) nucleon-particle configuration in \lam{10}{B}--\lam{10}{Be}. 
This enhancement follows from the relative magnitudes of the Fermi-like 
interaction term ${\bar V}^{0p}_{\Lambda\Sigma}$ and its Gamow-Teller partner 
term $\Delta^{0p}_{\Lambda\Sigma}$. Noting that both the $A=4$ and $A=8$ 
mirror hypernuclei correspond to SU(4) nucleon-hole configuration, the roughly 
factor two ratio of $\langle V_{\rm CSB}{\rangle}_{A=4}$=232~keV to $\langle 
V_{\rm CSB}{\rangle}_{A=8}$=119~keV reflects the approximate factor two ratio 
of $0s_N0s_Y$ to $0p_N0s_Y$ $\Lambda\Sigma$ matrix elements. However, in 
distinction from the $A$=4 g.s. isodoublet where $\Delta B_{\Lambda}\approx 
\langle V_{\rm CSB}\rangle$, the increasingly negative Coulomb contributions 
in the $p$-shell overcome the positive $\langle V_{\rm CSB}{\rangle}$ 
contributions, with $\Delta B_{\Lambda}$ becoming negative definite for 
$A\geq 9$. 

Comparing $\Delta B_{\Lambda}^{\rm calc}$ with $\Delta B_{\Lambda}^{\rm exp}$ 
in Table~\ref{tab:pshell}, we note the reasonable agreement reached between 
the $\Lambda\Sigma$ coupling model calculation and experiment for all five 
pairs of $p$-shell hypernuclei listed here. Extrapolating to heavier 
hypernuclei, one might naively expect negative values of $\Delta B_{\Lambda}^{
\rm calc}$. However, this assumes that the negative Coulomb contribution 
remains as large upon increasing $A$ as it is in the beginning of the $p$ 
shell, which need not be the case. As nuclear cores beyond $A=9$ become more 
tightly bound, the $\Lambda$ hyperon is unlikely to compress these nuclear 
cores as much as it does in lighter hypernuclei, so that the additional 
Coulomb repulsion in \lam{12}{C}, for example, over that in \lam{12}{B} may 
not be sufficiently large to offset the attractive CSB contribution to 
$B_{\Lambda}$(\lam{12}{C})$-B_{\Lambda}$(\lam{12}{B}), in agreement with the 
value 50$\pm$110~keV suggested recently for this $A$=12 $B_{\Lambda}$(g.s.) 
splitting using FINUDA and JLab counter measurements~\cite{botta17}. 
In making this argument one relies on the expectation, based on SU(4) 
supermultiplet fragmentation patterns in the $p$ shell, that $\langle 
V_{\rm CSB}\rangle$ does not exceed $\sim$100~keV. Based on $\Lambda\Sigma$ 
mixing model arguments given in Ref.~\cite{galmil13}, CSB splittings in 
medium-heavy and heavy hypernuclei become progressively negligible. 


\section{\lamb{3}{H} AND \lamb{3}{n} LIFETIME PUZZLES}
\label{sec:L3H} 

Measurements of the \lamb{3}{H} lifetime in emulsion or bubble-chamber 
experiments during the 1960s and early 1970s gave conflicting and puzzling 
results. Particularly troubling appeared a conference report by Block 
{\it et al.} claiming a lifetime of $\tau$(\lamb{3}{H})=(95$^{+19}_{-15}
$)~ps~\cite{Block62}, to be compared with a free $\Lambda$ lifetime 
$\tau_\Lambda$=(236$\pm$6)~ps measured in the same He chamber~\cite{Block63}. 
Given the loose $\Lambda$ binding, $B_\Lambda$(\lamb{3}{H})=0.13$\pm$0.05~MeV, 
it was anticipated that $\tau$(\lamb{3}{H})$\approx$$\tau_\Lambda$, 
as clearly seen in the Rayet and Dalitz~(RD)~\cite{RD66} approach in which 
the \lamb{3}{H} decay rate for any of its g.s. doublet members, whichever 
is the actual g.s., is given by 
\begin{equation} 
\Gamma_{_\Lambda^3{\rm H}}^{J=1/2}=\frac{\bar q}{1+\omega_{\pi}(\bar q)
/E_{3N}(\bar q)}[|s_{\pi}|^2(1+\frac{1}{2}\eta(\bar q))+|p_{\pi}|^2(
\frac{\bar q}{q_\Lambda})^2(1-\frac{5}{6}\eta(\bar q))],    
\label{eq:L3Hg.s.} 
\end{equation}  
\begin{equation} 
\Gamma_{_\Lambda^3{\rm H}}^{J=3/2}=\frac{\bar q}{1+\omega_{\pi}(\bar q)/E_{3N}
(\bar q)}[|s_{\pi}|^2(1-\eta(\bar q))+|p_{\pi}|^2(\frac{\bar q}{q_\Lambda})^2
(1-\frac{1}{3}\eta(\bar q))]. 
\label{eq:L3Hexc.} 
\end{equation} 
Here $\eta(\bar q)$, with values between 0 and 1, is an exchange integral that 
arises by assigning a fixed pion momentum $\bar q$ in a closure approximation 
to all pion-nuclear final states. It can be shown that $\eta(\bar q)\to 0$ 
when $B_\Lambda\to 0$, so apart from minor kinematical factors each 
\lamb{3}{H} decay rate reduces then to the free $\Lambda\to N+\pi$ decay rate 
\begin{equation} 
\Gamma_\Lambda(q_\Lambda)=\frac{q_\Lambda}{1+\omega_{\pi}(q_\Lambda)
/E_N(q_\Lambda)}(|s_{\pi}|^2+|p_{\pi}|^2),~~~~~ 
\left|\frac{p_{\pi}}{s_{\pi}}\right|^2 \approx 0.132, 
\label{eq:GammaLam} 
\end{equation} 
with $\omega_{\pi}(q_\Lambda)$ and $E_N(q_\Lambda)$ the center of mass 
energies of the decay pion and the recoil nucleon, respectively, and where 
$q_\Lambda\approx 102$~MeV/c is a weighted average for the $p+\pi^-$ and 
$n+\pi^0$ branches of the $\Lambda\to N + \pi$ weak decay. 

\begin{table}[htb] 
\caption{\lamb{3}{H}$_{\rm g.s.}({\frac{1}{2}}^+)$ decay rate calculated in 
units of the free $\Lambda$ decay rate $\Gamma_\Lambda$ and listed in a year 
of publication order. The first row lists results for plane-wave pions, 
disregarding pion final state interaction (FSI) contributions which are listed 
in the second row. A calculated nonmesonic decay rate contribution of 0.017 
from Ref.~\cite{Golak97} was added uniformly in obtaining the total decay 
rates listed in the third row. For a more critical assessment of the decay 
rates published in Refs.~\cite{RD66,Fad98}, see Ref.~\cite{GG18}.} 
\vspace{5pt}
{\renewcommand{\arraystretch}{1.1}}
\begin{tabular}{ccccc}
\hline 
$\Gamma$(\lamb{3}{H}) model & 1966~\cite{RD66} & 1992~\cite{Con92} & 
1998~\cite{Fad98} & 2019~\cite{GG18} \\  
\hline 
Without pion FSI & 1.05 & 1.12 & 1.01 & 1.11 \\ 
Pion FSI contribution & $-$0.013 & -- & -- & 0.11 \\  
Total & 1.05 & 1.14 & 1.03 & 1.23 \\  
\hline 
\end{tabular}
\label{tab:L3H}
\end{table}  

Table~\ref{tab:L3H} lists \lamb{3}{H}$_{\rm g.s.}({\frac{1}{2}}^+)$ decay 
rate values calculated by RD and in several subsequent calculations, 
all reaching similar results. Claims for large departures from the free 
$\Lambda$ value  are often found incorrect or irreproducible. The RD results 
were confirmed by Ram and Williams using a wider class of \lamb{3}{H} 
model wavefunctions~\cite{RW71}. The RD methodology was also used in the 
Congleton~\cite{Con92} and Gal-Garcilazo~(GG)~\cite{GG18} calculations, with 
the latter one solving appropriate three-body Faddeev equations to produce 
a \lamb{3}{H} wavefunction. The Kamada {\it et al.} calculation~\cite{Fad98}, 
while also solving Faddeev equations for the \lamb{3}{H} wavefunction, 
accounted microscopically for the outgoing 3$N$ phase space and FSI, thereby 
doing without a closure approximation. Pion FSI was considered only in two of 
these works, with differing results: (i) repulsion, weakly reducing $\Gamma
$(\lamb{3}{H}) in RD; and (ii) attraction, moderately enhancing it in GG. 
The latter result is supported by the $\pi^-$-atom $1s$ level {\it attractive} 
shift observed in $^3$He~\cite{Schwanner84}.  

\begin{figure}[!t]  
\includegraphics[width=0.9\textwidth]{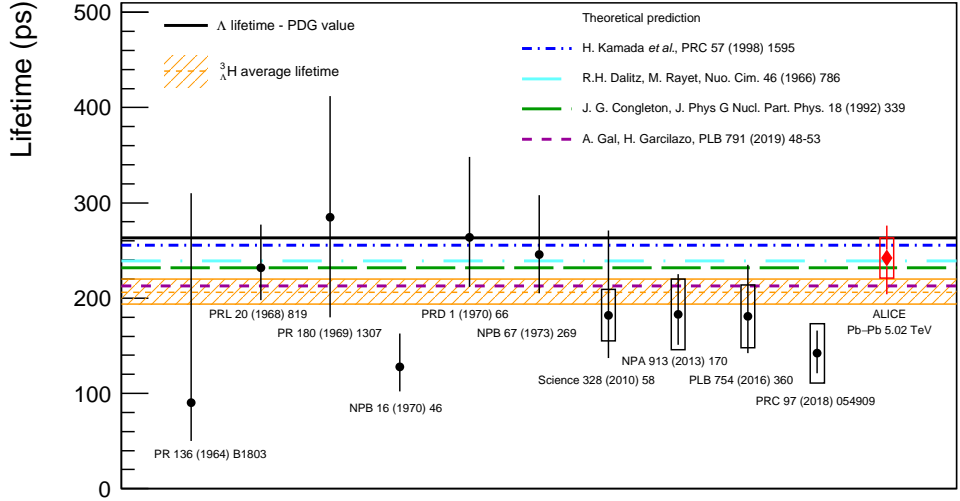} 
\caption{Compilation of measured \lamb{3}{H} lifetime values, 
plotted in chronological order. The five most recent values are from 
relativistic heavy ion experiments: STAR~\cite{STAR10}, HypHI~\cite{HypHI13}, 
ALICE~\cite{ALICE16}, STAR~\cite{STAR18} and ALICE~\cite{ALICE19}. Shown by 
horizontal lines are the free-$\Lambda$ lifetime (solid), the world average 
of measured \lamb{3}{H} lifetimes (dashed), and results from the four 
calculations listed in Table~\ref{tab:L3H}. Figure courtesy of Benjamin 
D\"{o}nigus~\cite{ALICE19}.} 
\label{fig:L3H} 
\end{figure} 

Renewed interest in the \lamb{3}{H} lifetime problem arose by recent 
measurements of $\tau$(\lamb{3}{H}) in relativistic heavy ion production 
experiments~\cite{BMD19} marked by rectangle systematic uncertainties in 
Fig.~\ref{fig:L3H}. Shown also is the world-average value which is shorter 
by about 30\% than the free-$\Lambda$ lifetime $\tau_\Lambda$=(263$\pm$2)~ps. 
In sharp contrast with the large scatter of old bubble chamber and nuclear 
emulsion measurements, these five recent measurements of $\tau$(\lamb{3}{H}) 
give values persistently shorter by (30$\pm$8)\% than $\tau_\Lambda$. Also 
shown in the figure are the $\tau_{\rm calc}$(\lamb{3}{H}) values listed 
in Table~\ref{tab:L3H}, all of which are insufficiently short to reproduce 
the recently measured shorter lifetime values, except the very recent one 
by ALICE~\cite{ALICE19}. While enhancement of the free $\Lambda$ decay rate 
by up to $\approx$20\% is theoretically conceivable~\cite{GG18}, it appears 
inconceivable to reproduce the 30\% or so enhancement suggested by the recent 
heavy-ion experiments. 

Before closing this section I wish to make a few remarks on \lamb{3}{n}, 
assuming it is bound as conjectured by the HypHI GSI 
Collaboration~\cite{Rap13} (but unanimously unbound in 
recent theoretical calculations~\cite{GV14,Hiyama14,GG14}). 
In \lamb{3}{n} decays induced by $\Lambda\to p+\pi^-$, where the \lamb{3}{n} 
neutrons are spectators, the \lamb{3}{n}~$\to (pnn) + \pi^-$ weak decay rate 
is given in the closure approximation essentially by the $\Lambda\to p+\pi^-$ 
free-space weak-decay rate, whereas in $\Lambda\to n+\pi^0$ induced decays 
the production of a third low-momentum neutron is suppressed by the Pauli 
principle, and this \lamb{3}{n} weak decay branch may be disregarded up to 
perhaps a few percents. We thus obtain~\cite{GG18}: 
\begin{equation} 
\Gamma_{_\Lambda^3{\rm n}}^{J=1/2}=\frac{\bar q}{1+\omega_{\pi}(\bar q)/E_{3N}
(\bar q)}\,0.641\,\left(|s_{\pi}|^2+|p_{\pi}|^2(\frac{\bar q}{q_{\Lambda}})^2
\right), 
\label{eq:L3n1} 
\end{equation} 
where the coefficient 0.641 is the free-space $\Lambda\to p+\pi^-$ fraction 
of the total $\Lambda\to N+\pi$ weak decay rate. Evaluating the ratio 
$\Gamma$(\lamb{3}{n})/$\Gamma_\Lambda$ for the choice ${\bar q}=q_\Lambda$ 
one obtains 
$\Gamma({_\Lambda^3}{\rm n})/\Gamma_\Lambda\approx 1.114\times 0.641=0.714$, 
where the factor 1.114 follows from the difference between 
$E_{3N}(q_\Lambda)$ and $E_{N}(q_\Lambda)$ in the phase space 
factors, giving rise to an estimated \lamb{3}{n} lifetime of 
$\tau$(\lamb{3}{n})$\,\approx 368$~ ps, 
which should hold up to a few percent contribution from the $\pi^0$ decay 
branch. This lifetime is considerably longer than 181${^{+30}_{-24}}\pm$25~ps 
or 190${^{+47}_{-35}}\pm$36~ps deduced from the $nd\pi^-$ and $t\pi^-$ alleged 
decay modes of \lamb{3}{n}~\cite{Rap13,Saito16}, providing a strong argument 
against the conjectured stability of \lamb{3}{n}.

\section{$\Lambda^\ast$(1405) MATTER?}  

\begin{figure}[!t]  
\includegraphics[width=0.6\textwidth]{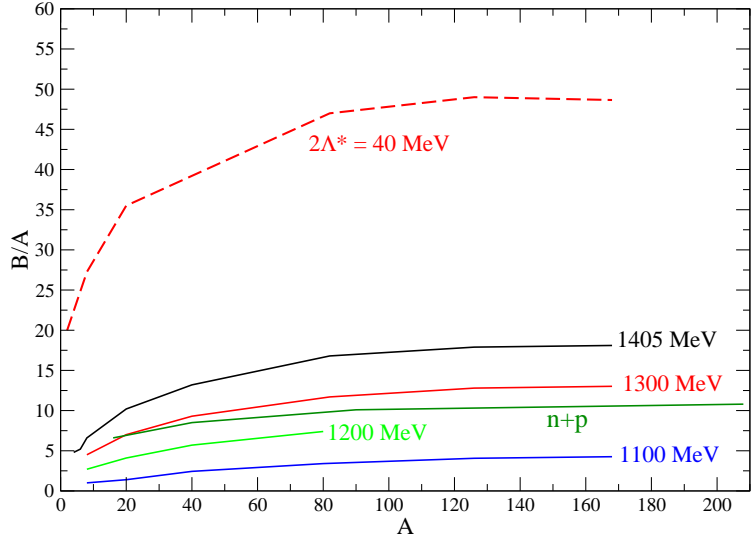} 
\caption{Binding energies per baryon, $B/A$, as a function of mass number 
$A$ in RMF calculations~\cite{plb18}. Solid curves use standard nuclear 
RMF meson-baryon coupling constants~\cite{hs81} for $n+p$ nuclei and for 
single-type baryon aggregates with variable mass as marked. The dashed curve 
for $\Lambda^{\ast}$ aggregates uses an increased scalar coupling constant 
such that $B(\Lambda^{\ast}\Lambda^{\ast})=40$~MeV, see Table~\ref{tab:KbarN}. 
Figure courtesy of J.~Hrt\'{a}nkov\'{a} and J.~Mare\v{s}.}
\label{fig:RMF} 
\end{figure} 

Recent Lattice QCD calculations~\cite{lqcd15} demonstrate that the $\Lambda^{
\ast}$(1405) $I=0$, $J^\pi = (1/2)^-$ hyperon is well described by a $\bar K 
N$ hadronic `molecule', as predicted in 1959 by Dalitz and Tuan~\cite{dt59}, 
rather than by a 3-quark baryon. The observed $\Lambda^\ast$(1405) structure 
requires a two-pole configuration dominated by a pole closer to the $\bar K N$ 
threshold than its nominal `1405' position 27 MeV below the $K^-p$ threshold 
might suggest~\cite{mh18}. The resulting $\bar K N$ effective potential is 
then necessarily {\it energy dependent}, as shown by recent applications of 
$\chi$EFT meson-baryon coupled-channel models that generate it dynamically 
\cite{mhw18}. The difference between this `energy-dependent' $\bar K N$ 
potential picture and the single-pole `energy-independent' $\bar K N$ 
potential picture advanced by Akaishi and Yamazaki~\cite{ay02,ya07} is seen in 
Table~\ref{tab:KbarN} to compound upon going to three- and four-body systems. 

The $\Lambda^\ast\Lambda^\ast$ binding energy deduced in Table~\ref{tab:KbarN} 
from the E-indep. calculations~\cite{may13} is exceptionally large. This was 
to be expected since the $\Lambda^\ast$ itself is a quite strongly bound $\bar 
K N$ state. It led Akaishi and Yamazaki to speculate, based apparently on 
non-documented nonrelativistic few-body calculations of $\Lambda^\ast$ 
aggregates, that absolute stability of $\Lambda^\ast$ configurations is 
achieved for $A\gtrsim 8$~\cite{plb17}. This speculation dates back to a work 
coauthored by them in 2004~\cite{yad04}. What these authors apparently missed 
was that solving the $A$-body Schr\"{o}dinger equation for purely attractive 
$\Lambda^{\ast}\Lambda^{\ast}$ interactions will necessarily lead to collapse, 
with the binding energy per $\Lambda^\ast$, $B/A$, and the central $\Lambda^{
\ast}$ density $\rho(r\approx 0)$ diverging as $A$ increases. The collapse 
of $\Lambda^\ast$ nuclei may be averted within relativistic mean field (RMF) 
calculations~\cite{plb18}, as briefly described below and further detailed by 
Hrt\'{a}nkov\'{a} in a separate contribution.

\begin{table}[htb] 
\caption{$(\bar K N)_{I=0}$, $(\bar K NN)_{I=1/2}$ and $(\bar K \bar K
NN)_{I=0}$ binding energies $B$ (in MeV) calculated using energy dependent
(E-dep.) and energy independent (E-indep.) $\bar K N$ potentials. 
$(\bar K \bar K NN)_{I=0}$ binding energies are transformed in the 
last column to $B_{\Lambda^{\ast}\Lambda^{\ast}}$ values. Table adapted 
from \cite{plb18}.} 
\vspace{5pt}
{\renewcommand{\arraystretch}{1.1}}
\begin{tabular}{ccccc}
\hline 
$\bar K$ model & $(\bar K N)_{I=0}$ & $(\bar K NN)_{I=1/2}$ & 
$(\bar K \bar K NN)_{I=0}$ & $\Lambda^{\ast}\Lambda^{\ast}$ \\ 
\hline 
(E-dep.)~\cite{bgl12} & 11.4 & 15.7 & 32.1 & 9.3 \\ 
(E-dep.)~\cite{dim18} & 11 & 14--28 & -- & -- \\ 
(E-indep.)~\cite{may13} & 26.6 & 51.5 & 93 & 40 \\ 
(E-indep.)~\cite{dim17} & 28 & 51 & -- & -- \\ 
\hline 
\end{tabular}
\label{tab:KbarN}
\end{table}  

Shown in Ref.~\cite{plb18}, within RMF calculations in which strongly 
attractive $\Lambda^{\ast}\Lambda^{\ast}$ interactions are generated through 
scalar meson ($\sigma$) and vector meson ($\omega$) exchanges, is that both 
$B/A$ and the central density $\rho(r\approx 0)$ saturate for values of $A$ 
of order $A\sim 100$: $B/A$ saturates at values between roughly 30 to 80 MeV, 
depending on details of the RMF modeling, and the associated central densities 
saturate at values about twice nuclear-matter density. A simple example 
of $B/A$ saturation is shown in Fig.~\ref{fig:RMF} where the calculated 
$\Lambda^{\ast}$-aggregates binding energies (dashed curve) are normalized 
to yield $B(\Lambda^{\ast}\Lambda^{\ast})=40$~MeV from Table~\ref{tab:KbarN}. 
In these RMF calculations, done for $\Lambda^\ast$ closed-shell nuclei, 
one solves self consistently a coupled system of Klein-Gordon equations for 
the meson fields and a Dirac equation for $\Lambda^\ast$ that result from a 
$\Lambda^{\ast}$ RMF lagrangian density given by
\begin{equation}  
\mathcal L = \bar{\Lambda}^* \left[\,{\rm i}\gamma^\mu D_\mu-(M_{\Lambda^*}-
g_{\sigma \Lambda^*} \sigma) \right] \Lambda^* + (\sigma, \omega_\mu\,
\textrm{free-field terms})~, 
\label{eq:Lag2} 
\end{equation} 
where $D_\mu=\partial_\mu+{\rm i}\,g_{\omega \Lambda^*}\,\omega_\mu$, 
$M_{\Lambda^*}$ is the mass of $\Lambda^*$ and $g_{i \Lambda^*}$ 
($i=\sigma, \omega$) are the corresponding coupling constants. 


With calculated values of $B/A < 100$~MeV, $\Lambda^{\ast}$ aggregates remain 
highly unstable against strong-interaction decay governed by two-body 
conversion reactions such as $\Lambda^{\ast}\Lambda^{\ast}\to\Lambda\Lambda, 
\Sigma\Sigma$. This preserves the strong-interaction stable form of hadronic 
matter, known as `strange hadronic matter', in which SU(3)-octet hyperons 
($\Lambda,\Sigma,\Xi$) exclusively are as abundant as nucleons \cite{Schaf}. 
Considering that $\Lambda^{\ast}$s stand for $\bar K N$ bound states, this 
conclusion was in fact reached already 10 years ago by Gazda {\it et al.} in 
RMF calculations of multi-$\bar K$ nuclei, where for a given core nucleus the 
resulting $\bar K$ separation energies $B_{\bar K}$, as well as the associated 
nuclear and $\bar K$-meson densities, saturate with the number $\kappa$ of 
$\bar K$ mesons for $\kappa \geq 10$~\cite{gfgm08}. In these calculations 
$B_{\bar K}$ generally does not exceed 200 MeV which is insufficient to 
compete with multi-hyperonic nuclei in providing the ground state of strange 
hadronic configurations; put differently, it means that kaon condensation is 
unlikely to occur in strong-interaction self-bound strange hadronic matter. 

The RMF calculations discussed above demonstrate the decisive role of 
Lorentz covariance in producing saturation of binding energies and sizes. 
Lorentz covariance introduces two types of baryon density, a scalar 
$\rho_{\rm S}={\bar B}B$ associated with the attractive $\sigma$ meson field 
and a vector $\rho_{\rm V}={\bar B}\gamma_0 B$ associated with the repulsive 
$\omega$ meson field. Whereas $\rho_{\rm V}$ coincides with the conserved 
baryon density $B^\dag B$, $\rho_{\rm S}$ shrinks with respect to 
$\rho_{\rm V}$ in dense matter by a multiplicative factor $M^\ast/E^\ast 
<1$, where $M^\ast=M-g_{\sigma B}\,\langle\sigma\rangle < M$ is the baryon 
density-dependent effective mass, thereby damping the attraction from the 
scalar $\sigma$ meson field and maintainig saturation. In contrast, 
non-relativistic calculations with static potentials that underlie the 
standard RMF nuclear calculations would lead to collapse of systems 
composed of sufficiently large number of $\Lambda^\ast$ baryons, as it 
also holds for nucleons (M. Sch\"{a}fer, private communication).

\section{SUMMARY}

Here I comment briefly on three of the four topics discussed in earlier 
sections.
\newline 
\newline 
\noindent
\textbf{Overbinding in $\Lambda$ hypernuclei}.~~The discussion above made 
it clear that $\Lambda NN$ repulsive interactions are required to resolve 
the overbinding problem of \lamb{5}{He}. One may ask whether these $\Lambda 
NN$ interaction terms are of similar magnitude to those used by the QMC 
practitioners across the periodic table, namely $\Lambda NN$~(I,II) in 
Fig.~\ref{fig:lpg}. The figure suggests that their two-body attractive 
$\Lambda N$ interaction is exceedingly strong, perhaps leading on its own to 
a $\Lambda$ well depth of order $D^{(2)}_\Lambda\gtrsim 100$~MeV. If so, the 
compensating $\Lambda NN$ repulsive interaction is also huge, canceling most 
of the $\Lambda N$ attraction in reaching the phenomenologically deduced value 
of $D_\Lambda\sim 30$~MeV. To estimate $D^{(2)}_\Lambda$, one starts from 
the 1st order $\Lambda$-nucleus optical potential `$t\rho$' form, including 
long-range Pauli correlations~\cite{wrw97}, 
\begin{equation} 
V_\Lambda(\rho)= -\frac{4\pi}{2m_\Lambda}a^{\rm lab}_{\Lambda N}(\rho)\rho,~~
~~a^{\rm lab}_{\Lambda N}(\rho)=\frac{a^{\rm lab}_{\Lambda N}}
{1+(3/2\pi)k_F(\rho)a^{\rm lab}_{\Lambda N}}\,, 
\label{eq:Vopt} 
\end{equation}
in terms of a nuclear-medium Pauli corrected $\Lambda N$ scattering length 
$a^{\rm lab}_{\Lambda N}(\rho)$ and nuclear density $\rho=2k_F^3/3\pi^2$ 
where $k_F$ is the local Fermi momentum. It was assumed that the singlet 
and triplet $\Lambda N$ scattering lengths assume the same value which is 
taken directly from from the Alexander {\it et al.} $\Lambda p$ scattering 
experiment~\cite{Alex68}: $a^{\rm cm}_{\Lambda N}=1.65$~fm. For nuclear-matter 
density $\rho_0=0.170$~fm$^{-3}$ corresponding to Fermi momentum $k_F=1.36
$~fm$^{-1}$, one obtains $D^{(2)}_\Lambda = -V_\Lambda(\rho) = 40.2$~MeV 
which roughly in the range of values reached using Nijmegen soft-core and 
extended soft core potentials~\cite{sr13}, but much smaller than the value 
reached by the QMC practitioners. This observation requires further studies. 
\newline 
\newline 
\noindent
\textbf{CSB in hypernuclei}.~~It was shown above that the Gazda-Gal 
calculations~\cite{gg16a,gg16b} of the \lamb{4}{He}--\lamb{4}{H} binding 
energy difference in the $0^+$ g.s. were able to reproduce the order of 
magnitude around 200~keV of this CSB splitting using a simple prescription, 
Eq.~(\ref{eq:OME}), suggested by $\Lambda-\Sigma^0$ mixing. This mixing 
induces a long-range OPE CSB interaction related via (\ref{eq:OME}) to the 
strong-interaction OPE $\Lambda\Sigma$ coupling. Other hadron mixings, 
$\omega -\rho^0$ and $\eta -\pi^0$, generate considerably shorter-range CSB 
potentials that apparently are much weaker that the OPE potential induced by 
$\Lambda-\Sigma^0$ mixing. In the $A$=3 core nuclei, 
the $^3$H--$^3$He binding energy difference of 764~keV is dominated by the 
Coulomb interaction, leaving about 70~keV to CSB which is believed to arise 
mostly from $\omega - \rho^0$ mixing~\cite{Miller06}. Considering that only 
$\omega$, but not $\rho^0$ couples to $\Lambda$, and that the $\omega\Lambda$ 
coupling constant is roughly 2/3, in SU(6), of the $\omega N$ coupling 
constant, the $\omega - \rho^0$ mixing effect in the $A$=4 hypernuclei should 
be roughly 1/3 of its contribution in in the $A$=3 rnuclei, namely about 20 
keV in size. Indeed, this is close to the $\omega -\rho^0$ mixing contribution 
in the $A$=4 hypernuclei reported by Coon {\it et al.}~\cite{Coon98} who also 
found even a smaller $\eta - \pi^0$ contribution. In conclusion, it appears 
that the CSB interaction operative in $\Lambda$ hypernuclei is dominated by 
the OPE component generated by $\Lambda-\Sigma^0$ mixing. 
\newline 
\newline 
\noindent
\textbf{Lifetimes of light hypernuclei}.~~It was shown that a lifetime 
of \lamb{3}{H} shorter by up to $\approx$20\% than the free $\Lambda$ 
lifetime may be accommodated by theory. A new theoretical development is 
the $\sim$10\% estimated contribution from the attractive pion FSI for 
$A$=3~\cite{GG18}. If confirmed, and perhaps found even somewhat larger, 
it might bridge the gap between our present $\approx$20\% theoretical estimate 
for the reduction of $\tau$(\lamb{3}{H}) with respect to $\tau_\Lambda$ and 
the $\approx$30\% reported in relativistic heavy-ion collision experiments. 
It is remarkable that \lamb{3}{H} decay is the only light hypernucleus decay 
where the pion FSI is expected to be attractive. The decays of \lamb{4}{H}, 
\lamb{4}{He} and \lamb{5}{He} involve pion-$^4$He FSI which is known from the 
$\pi^-$ atomic $^4$He $1s$ level shift to be repulsive~\cite{Backenstoss74}. 
Disregarding such pion FSI in the decay of the $A$=4 hypernuclei, simple 
estimates of their lifetimes agree well with the KEK measured 
values~\cite{Outa98}:  
\begin{equation} 
\tau(_\Lambda^4{\rm H})=194^{+24}_{-26}~{\rm ps}, ~~~~~ 
\tau(_\Lambda^4{\rm He})=256\pm 27~{\rm ps}. 
\label{eq:tau4} 
\end{equation} 
The difference between these lifetimes arises mostly from the difference 
between the strongest partial decays which are the two-body decays to 
$\pi$+$^4$He, with a rate calculated reliably as $\approx 0.7\Gamma_\Lambda$ 
for \lamb{4}{H} and half of that for \lamb{4}{He}, reflecting the $\pi^-$ to 
$\pi^0$ weights in the free $\Lambda$ decay. Hence the difference between the 
\lamb{4}{H} and \lamb{4}{He} total decay rates should amount to $\approx 0.35
\Gamma_\Lambda$, close to the $\approx 0.33\Gamma_\Lambda$ given by the 
measured lifetimes (\ref{eq:tau4}).

\section{ACKNOWLEDGMENTS}
Many thanks are due to Liguang Tang and his JLab organizing team for the 
gracious hospitality bestowed on me during the HYP2018 International 
Conference at Portsmouth-Norfolk, VA. I am also indebted to my good colleagues 
Nir Barnea, Lorenzo Contessi, Eli Friedman, Humberto Garcilazo, Daniel Gazda, 
Jaroslava Hrt\'{a}nkov\'{a}, Ji\v{r}\'{i} Mare\v{s}, John Millener, Moreh 
Nevuchim, and Martin Sch\"{a}fer for many insights gained through our 
long-standing collaborations.

\end{document}